\documentclass[twocolumn,english,aps,prl,floatfix,superscriptaddress,showpacs,amsmath,amssymb]{revtex4}
\usepackage[utf8]{luainputenc}
\usepackage{amsmath}
\usepackage{amssymb}
\usepackage{stackrel}
\usepackage{graphicx}
\usepackage{esint}

\makeatletter
\@ifundefined{textcolor}{}
{%
 \definecolor{BLACK}{gray}{0}
 \definecolor{WHITE}{gray}{1}
 \definecolor{RED}{rgb}{1,0,0}
 \definecolor{GREEN}{rgb}{0,1,0}
 \definecolor{BLUE}{rgb}{0,0,1}
 \definecolor{CYAN}{cmyk}{1,0,0,0}
 \definecolor{MAGENTA}{cmyk}{0,1,0,0}
 \definecolor{YELLOW}{cmyk}{0,0,1,0}
}

\usepackage[normalem]{ulem}

\usepackage{babel}

\usepackage{babel}

\usepackage{babel}

\makeatother

\usepackage{babel}
\begin{document}

\title{d-wave superconductivity in boson+fermion dimer models}

\author{Garry Goldstein}

\affiliation{TCM Group, Cavendish Laboratory, University of Cambridge, J. J. Thomson
Avenue, Cambridge CB3 0HE, United Kingdom}

\author{Claudio Chamon}

\affiliation{Department of Physics, Boston University, Boston, Massachusetts 02215,
USA}

\author{Claudio Castelnovo}

\affiliation{TCM Group, Cavendish Laboratory, University of Cambridge, J. J. Thomson
Avenue, Cambridge CB3 0HE, United Kingdom}
\begin{abstract}
We present a slave-particle mean-field study of the mixed boson+fermion
quantum dimer model introduced by Punk, Allais, and Sachdev {[}PNAS~\textbf{112},
9552 (2015){]} to describe the physics of the pseudogap phase in cuprate
superconductors. Our analysis naturally leads to four charge $e$
fermion pockets whose total area is equal to the hole doping $p$,
for a range of parameters consistent with the $t-J$ model for high
temperature superconductivity. Here we find that the dimers are unstable
to d-wave superconductivity at low temperatures. The region of the
phase diagram with d-wave rather than s-wave superconductivity matches
well with the appearance of the four fermion pockets. In the superconducting
regime, the dispersion contains eight Dirac cones along the diagonals
of the Brillouin zone. 
\end{abstract}
\maketitle
\textit{\label{Introduction}Introduction.} The Rokhsar-Kivelson quantum
dimer model (QDM) was originally introduced to describe a possible
magnetically-disordered phase -- the resonating valence bond (RVB)
phase -- in high-temperature superconducting materials~\cite{key-1}.
The arena where the QDM has been deployed has greatly expanded since
its inception, and the model has taken on a key role in the study
of a variety of magnetic quantum systems. Quantum dimers show up prominently
in the study of hard-core bosons hopping on frustrated lattices~\cite{key-2},
of arrays of Josephson junctions with capacitative and Josephson couplings
\cite{key-3}, of frustrated Ising models with an external field or
with perturbative XY couplings~\cite{key-4}, of various types of
gauge theories~\cite{key-5}, and of models with large spin-orbit
couplings~\cite{key-6} and various cold atom setups~\cite{key-7}.
The study of QDMs led to an abundance of new phenomena including deconfined
quantum criticality and new routes to deconfinement~\cite{key-8}.
It also provided one of the earliest known examples of topologically
ordered states in a lattice model~\cite{key-25}.

\begin{figure}
\begin{centering}
\hspace{-0.5cm}\includegraphics[scale=0.13]{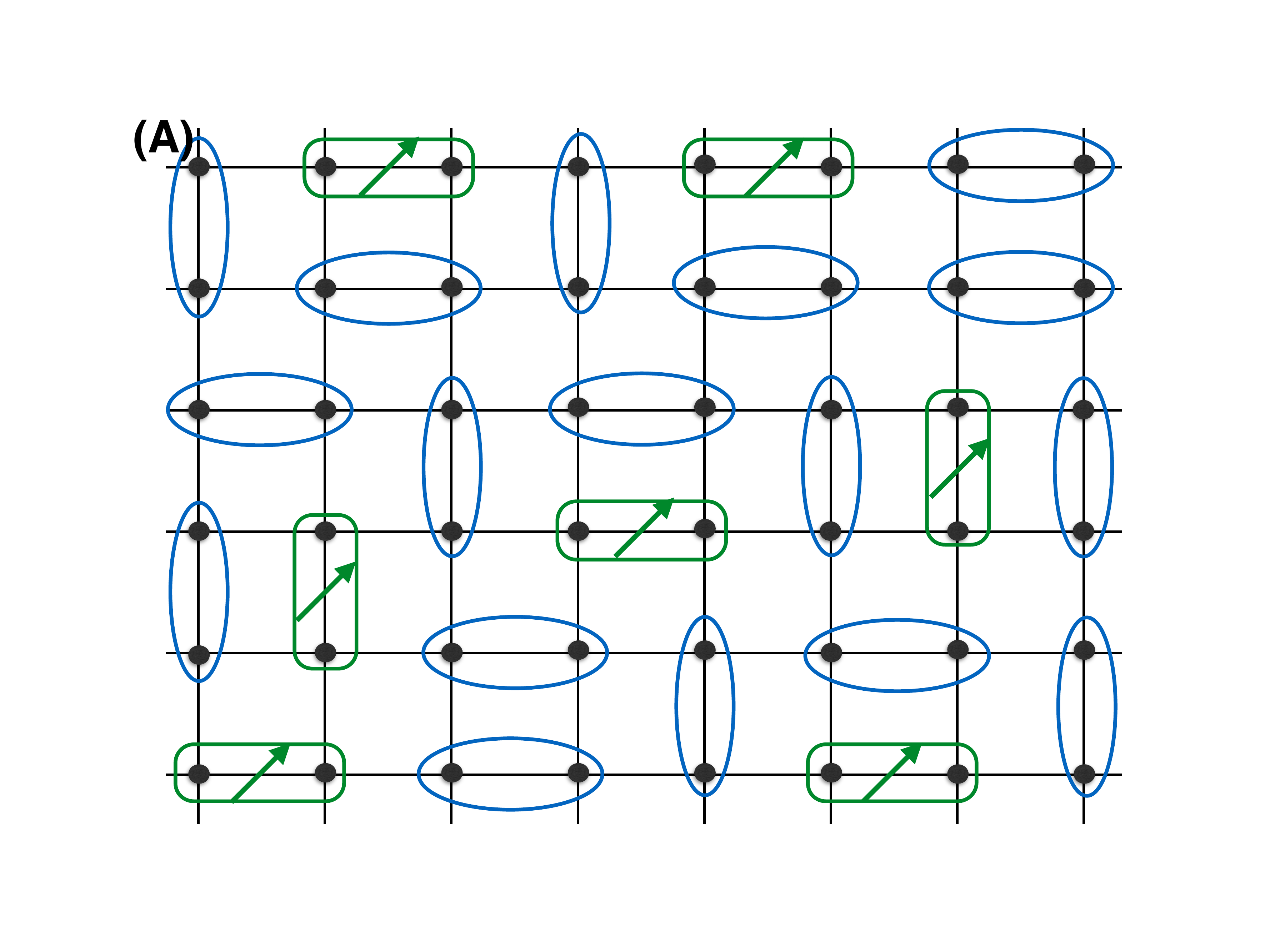}\hspace{-0.5cm}\includegraphics[scale=0.13]{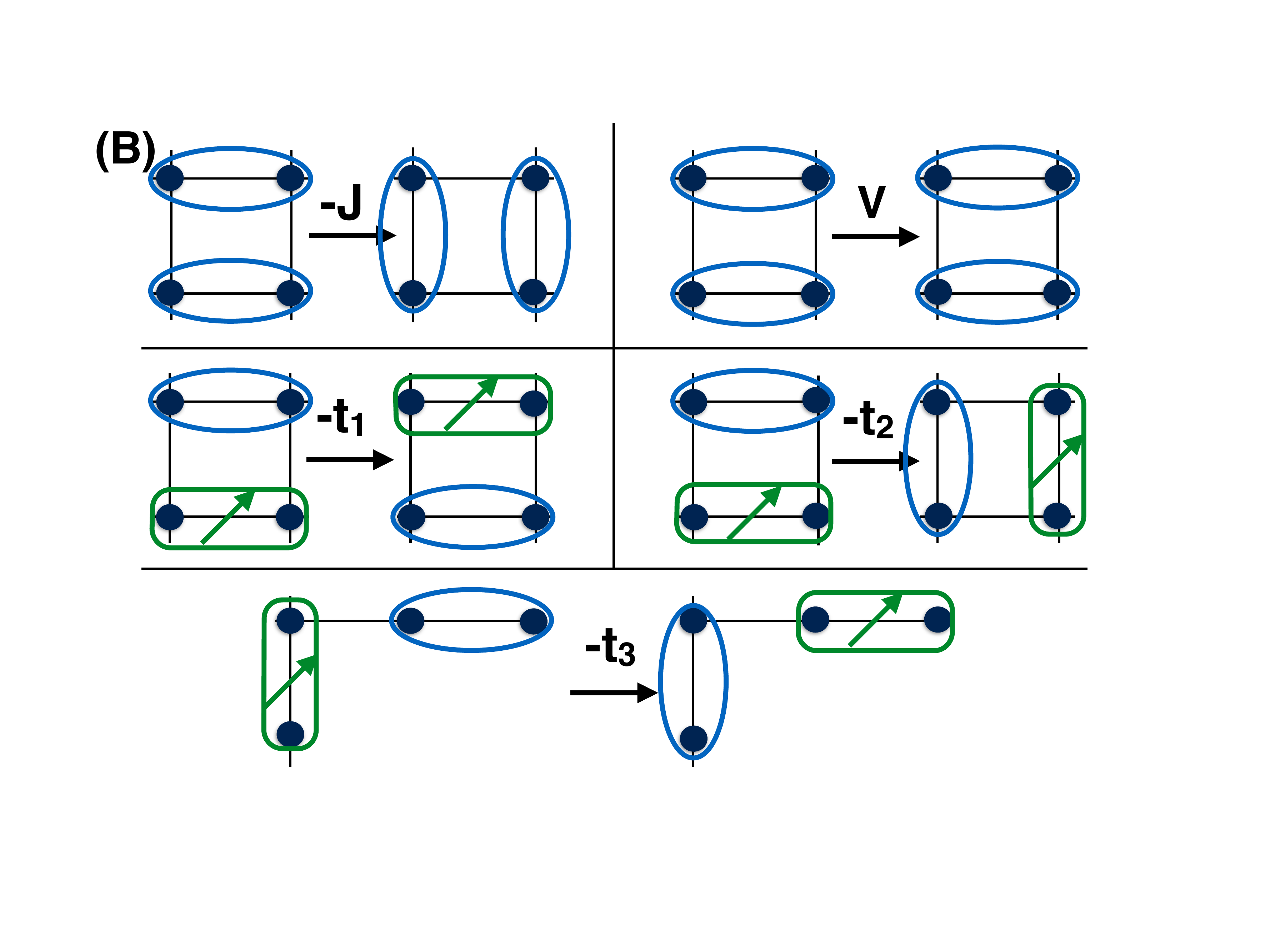} 
\par\end{centering}

\protect\protect\vspace{-0.7cm}
 \protect\protect\protect\protect\protect\caption{\label{fig:configurations} The boson+fermion quantum dimer model
of Ref.~\onlinecite{key-30}. (A) A particular dimer configuration.
The lattice is shown in black. The bosonic dimers representing the
valence bonds are shown in blue while the spinful fermionic dimers
representing a single electron delocalized over two sites are shown
in green. (B) Diagrams representative of the various terms in the
dimer Hamiltonian Eq.~\eqref{eq:Hamiltonian_Total}.}
\end{figure}

Recently QDMs have been revisited as models of high-temperature superconductivity
\cite{key-30,key-32,key-34}. This was motivated by the need to reconcile
transport experiments~\cite{key-44,key-45,key-46,key-29} and photoemission
data~\cite{key-41,key-42,key-43} in the underdoped region of cuprate
superconductors: while photoemission data show Fermi arcs enclosing
an area $1+p$ (with $p$ being the doping), transport measurements
indicate plain Fermi-liquid properties consistent with an area $p$.
In order to resolve this issue and produce a Fermi liquid which encloses
an area $p$, the authors of Refs.~\cite{key-30,key-32,key-34} introduced
a model for the pseudogap region of the cuprate superconductors which
consists of two types of dimers: one spinless bosonic dimer -- representing
a valence bond between two neighboring spins -- and one spin $1/2$
fermionic dimer representing a hole delocalized between two sites.
Fig.~\ref{fig:configurations} shows an example of a boson+fermion
dimer covering of the square lattice and depicts the dimer moves dictated
by the quantum Hamiltonian in Eq.~\eqref{eq:Hamiltonian_Total}.
The boson+fermion QDM (bfQDM) was introduced and studied numerically
in Ref.~\onlinecite{key-30} using exact diagonalization, supporting
the existence of a fractionalized Fermi liquid enclosing an area $p$.

In this work we present a slave boson and fermion formulation of the
bfQDM. We find that four symmetric fermion pockets, located in the
vicinity of $\left(\pm\frac{\pi}{2},\pm\frac{\pi}{2}\right)$ in the
Brillouin zone, naturally appear at mean-field level. The total area
of the four pockets is given by the hole (fermionic) doping. We find
that the system is unstable to d-wave superconductivity at low temperatures.
The region of the phase diagram with d-wave superconductivity matches
well the region with four fermion pockets. In the superconducting
phase, the fermionic dimers (holes) acquire a Dirac dispersion at
eight points along the diagonals of the Brillouin zone. 

\label{sec:Mapping-onto-slave} \textit{Mapping onto slave boson/fermion
model.} The quantum dimer model can be mapped exactly onto a slave
boson+fermion model by considering a secondary Hilbert space where
we assign to each link ($i,i+\eta$) of the lattice ($\eta=\hat{x},\hat{y}$)
a bosonic mode $b_{i,\eta}$ and a spinful fermionic mode $c_{i,\eta,\sigma}$
($\sigma=\uparrow,\downarrow$). We associate the number of dimers
on a link with the occupation numbers of the bosons or fermions on
that link. As such we have embedded the dimer Hilbert space in a larger
boson/fermion Hilbert space. The constraint that each site of the
lattice has one and exactly one dimer attached to it may be rephrased
in the boson/fermion language as: 
\begin{equation}
\Pi_{i}\equiv\sum_{l\in v_{i}}b_{l}^{\dagger}b_{l}^{\;}+c_{l,\uparrow}^{\dagger}c_{l,\uparrow}^{\;}+c_{l,\downarrow}^{\dagger}c_{l,\downarrow}^{\;}-1=0\,.\label{eq:Constraint}
\end{equation}
Here, for convenience of notation, $\ell\in v_{i}$ labels the four
links $j,\eta$ that are attached to vertex $i$. Any Hamiltonian
for the dimers has a boson/fermion representation; in particular the
terms illustrated in Fig.~\ref{fig:configurations}B can be written
as: 
\begin{align}
H_{D}= & \sum_{i}\left\{ -J\,b_{i,\hat{x}}^{\dagger}b_{i+\hat{y},\hat{x}}^{\dagger}b_{i,\hat{y}}^{\;}b_{i+\hat{x},\hat{y}}^{\;}+1\:\text{term}\right.\nonumber \\
 & \left.+V\,b_{i,\hat{x}}^{\dagger}b_{i,\hat{x}}^{\;}b_{i+\hat{y},\hat{x}}^{\dagger}b_{i+\hat{y},\hat{x}}^{\;}+1\:\text{term}\right\} \nonumber \\
 & \sum_{i}\sum_{\sigma}\left\{ -t_{1}\,b_{i,\hat{x}}^{\dagger}c_{i+\hat{y},\hat{x},\sigma}^{\dagger}c_{i,\hat{x},\sigma}^{\;}b_{i+\hat{y},\hat{x}}^{\;}+3\:\text{terms}\right.\nonumber \\
 & -t_{2}\,b_{i+\hat{x},\hat{y}}^{\dagger}c_{i,\hat{y},\sigma}^{\dagger}c_{i,\hat{x},\sigma}^{\;}b_{i+\hat{y},\hat{x}}^{\;}+7\:\text{terms}\nonumber \\
 & -t_{3}\,b_{i+\hat{x}+\hat{y},\hat{x}}^{\dagger}c_{i,\hat{y},\sigma}^{\dagger}c_{i+\hat{x}+\hat{y},\hat{x},\sigma}^{\;}b_{i,\hat{y}}^{\;}+7\:\text{terms}\nonumber \\
 & \left.-t_{3}\,b_{i+2\hat{y},\hat{x}}^{\dagger}c_{i,\hat{y},\sigma}^{\dagger}c_{i+2\hat{y},\hat{x},\sigma}^{\;}b_{i,\hat{y}}^{\;}+7\:\text{terms}\right\} \nonumber \\
 & -\mu\sum_{i}\sum_{\sigma}\sum_{\eta}c_{i,\eta,\sigma}^{\dagger}c_{i,\eta,\sigma}^{\;}\,,\label{eq:Hamiltonian_Total}
\end{align}
where we included a chemical potential for the holes (fermionic dimers),
which is important for the connection with doped high-temperature
superconductors~\cite{key-40,key-48}. The terms not written explicitly
in Eq.~\eqref{eq:Hamiltonian_Total} are simply obtained from those
shown by translational symmetry, four fold rotational symmetry, and
reflection symmetry about the two diagonals. This Hamiltonian also
has a local $U\left(1\right)$ gauge symmetry 
\begin{equation}
b_{i,\eta}\rightarrow e^{i\theta_{i}}\;b_{i,\eta}\;e^{i\theta_{i+\eta}}\;,\quad c_{i,\eta,\sigma}\rightarrow e^{i\theta_{i}}\;c_{i,\eta,\sigma}\;e^{i\theta_{i+\eta}}\;,\label{eq:Gauge_transformation}
\end{equation}
with a phase $\theta_{i}$ associate to each vertex $i$. Any Hamiltonian
that preserves the constraint given in Eq.~\eqref{eq:Constraint}
is invariant under this gauge transformation~\cite{key-31,key-33}.

A slave boson/fermion formulation of the bfQDM is obtained by introducing
a Lagrange multiplier: a real field $\lambda_{i}(\tau)$ that enforces
the dimer constraint Eq.~\eqref{eq:Constraint} at all times $\tau$,
and shifting the action by $\Delta{\cal S}=-\int d\tau\sum_{i}\lambda_{i}(\tau)\,\Pi_{i}(\tau)$.

\label{sec:A-variety-of}\textit{Slave boson/fermion mean-field decoupling.}
A systematic mean-field approach can be obtained by taking the saddle
point with respect to the Lagrange multiplier field $\lambda_{i}(\tau)\to\lambda_{i}$,
with a time-independent value $\lambda_{i}$ that enforces the average
constraint $\langle\Pi_{i}\rangle=0$. This procedure is accompanied
by Hubbard Stratonovich (HS) transformations of every term in the
Hamiltonian in Eq.~\eqref{eq:Hamiltonian_Total} separately. We begin
with the purely bosonic potential term: 
\begin{align}
 & b_{i,\hat{x}}^{\dagger}b_{i,\hat{x}}^{\;}b_{i+\hat{y},\hat{x}}^{\dagger}b_{i+\hat{y},\hat{x}}^{\;}\rightarrow\nonumber \\
 & \:\kappa_{1}\left\{ b_{i,\hat{x}}^{\dagger}b_{i,\hat{x}}^{\;}\,x_{i1}+b_{i+\hat{y},\hat{x}}^{\dagger}b_{i+\hat{y},\hat{x}}^{\;}\,x_{i2}-x_{i1}x_{i2}\right\} \nonumber \\
 & +\left(1-\kappa_{1}\right)\left\{ b_{i,\hat{x}}^{\dagger}b_{i+\hat{y},\hat{x}}^{\dagger}\,z_{i}+b_{i,\hat{x}}^{\;}b_{i+\hat{y},\hat{x}}^{\;}\,z_{i}^{*}-\left|z_{i}^{2}\right|\right\} ,\label{eq:Hubbard_Stratonovich_potential}
\end{align}
where $x_{i}$ and $z_{i}$ are auxiliary fields to be integrated
over and $\kappa_{1}$ is arbitrary. At mean-field level we can drop
the integrals over the auxiliary fields and replace them with their
saddle point values $x_{i1}\rightarrow\langle b_{i+\hat{y},\hat{x}}^{\dagger}b_{i+\hat{y},\hat{x}}^{\;}\rangle$,
$x_{i2}\rightarrow\langle b_{i,\hat{x}}^{\dagger}b_{i,\hat{x}}^{\;}\rangle$
and $z_{i}\rightarrow\langle b_{i,\hat{x}}^{\;}b_{i+\hat{y},\hat{x}}^{\;}\rangle$.
The hopping term may be decoupled in a similar manner: 
\begin{align}
 & b_{i,\hat{x}}^{\dagger}b_{i+\hat{y},\hat{x}}^{\dagger}b_{i,\hat{y}}^{\;}b_{i+\hat{x},\hat{y}}^{\;}+\text{h.c.}\rightarrow\nonumber \\
 & \:\kappa_{2}\left\{ b_{i,\hat{x}}^{\dagger}b_{i+\hat{y},\hat{x}}^{\dagger}\,w_{i1}+b_{i,\hat{y}}^{\;}b_{i+\hat{x},\hat{y}}^{\;}\,w_{i2}^{*}-w_{i1}w_{i2}^{*}+\text{h.c.}\right\} \nonumber \\
 & +\left(1-\kappa_{2}\right)\left\{ b_{i,\hat{x}}^{\dagger}b_{i,\hat{y}}^{\;}\,q_{i1}+b_{i+\hat{y},\hat{x}}^{\dagger}b_{i+\hat{x},\hat{y}}^{\;}\,q_{i2}^{*}-q_{i1}q_{i2}^{*}+\text{h.c.}\right\} ,\label{eq:Hubbard_Stratonovich_kinetic}
\end{align}
where, again, at mean-field level we use the saddle point values $w_{i1}\rightarrow\langle b_{i,\hat{y}}^{\;}b_{i+\hat{x},\hat{y}}^{\;}\rangle$,
$w_{i2}^{*}\rightarrow\langle b_{i,\hat{x}}^{\dagger}b_{i+\hat{y},\hat{x}}^{\dagger}\rangle$,
$q_{i1}\rightarrow\langle b_{i+\hat{y},\hat{x}}^{\dagger}b_{i+\hat{x}^{\;},\hat{y}}\rangle$,
$q_{i2}^{*}\rightarrow\langle b_{i,\hat{x}}^{\dagger}b_{i,\hat{y}}\rangle$
and $\kappa_{2}$ is arbitrary. Other HS decouplings, and linear combinations
thereof, are also possible.

We can make substantial progress in understanding the fermionic component
of the theory without detailed analysis of the bosonic component.
Indeed, any translationally invariant (liquid-like) bosonic ansatz,
naturally expected in $U\left(1\right)$ gauge theories coupled to
fermions with a Fermi surface~\cite{key-37,key-38,key-39}, yields
similar fermionic effective theories. The fermionic mean-field Hamiltonian
reads 
\begin{align}
H_{F\bar{B}} & =\sum_{\sigma}\sum_{i}\left\{ -t_{1}\;c_{i+\hat{y},\hat{x},\sigma}^{\dagger}c_{i,\hat{x},\sigma}^{\;}\langle b_{i,\hat{x}}^{\dagger}b_{i+\hat{y},\hat{x}}^{\;}\rangle+3\;\text{terms}\right.\nonumber \\
 & -t_{2}\;c_{i,\hat{y},\sigma}^{\dagger}c_{i,\hat{x},\sigma}^{\;}\langle b_{i+\hat{x},\hat{y}}^{\dagger}b_{i+\hat{y},\hat{x}}^{\;}\rangle+7\;\text{terms}\nonumber \\
 & -t_{3}\;c_{i,\hat{y},\sigma}^{\dagger}c_{i+\hat{x}+\hat{y},\hat{x},\sigma}^{\;}\langle b_{i+\hat{x}+\hat{y},\hat{x}}^{\dagger}b_{i,\hat{y}}^{\;}\rangle+7\;\text{terms}\nonumber \\
 & \left.-t_{3}\;c_{i,\hat{y},\sigma}^{\dagger}c_{i+2\hat{y},\hat{x},\sigma}^{\;}\langle b_{i+2\hat{y},\hat{x}}^{\dagger}b_{i,\hat{y}}^{\;}\rangle+7\;\text{terms}\right\} \nonumber \\
 & \left(-2\lambda-\mu\right)\sum_{i}\sum_{\sigma}\sum_{\eta}c_{i,\eta,\sigma}^{\dagger}c_{i,\eta,\sigma}^{\;}\;,\label{eq:Hamiltonian_boson_fermion_meanfield}
\end{align}
which is effectively a tight-biding model with renormalized hoppings
$T_{1}=t_{1}\;\langle b_{i,\hat{x}}^{\dagger}b_{i+\hat{y},\hat{x}}^{\;}\rangle$,
$T_{2}=t_{2}\;\langle b_{i+\hat{x},\hat{y}}^{\dagger}b_{i+\hat{y},\hat{x}}^{\;}\rangle$
and $T_{3}=t_{3}\;\langle b_{i+\hat{x}+\hat{y},\hat{x}}^{\dagger}b_{i,\hat{y}}^{\;}\rangle$.

The resulting model is defined on the bipartite checkerboard lattice
that is medial to the original square lattice. The horizontal ($x$)
and vertical ($y$) links make up the two sublattices where the fermions
reside. We define (in momentum space) the spinor that encodes these
two flavors as $\psi_{\vec{k},\sigma}^{\dagger}=(c_{\vec{k},\hat{y},\sigma}^{\dagger},c_{\vec{k},\hat{x},\sigma}^{\dagger})$
and 
\begin{equation}
H_{F\bar{B}}=\sum_{\vec{k},\sigma}\psi_{\vec{k},\sigma}^{\dagger}\,
\begin{pmatrix}\xi_{\vec{k}}^{x} & \gamma_{\vec{k}}\\
\gamma_{\vec{k}}^{*} & \xi_{\vec{k}}^{y}
\end{pmatrix}\;\psi_{\vec{k},\sigma}\;,\label{eq:H_k}
\end{equation}
where: 
\begin{align*}
\xi_{\vec{k}}^{x} & =-2\lambda-\mu-2\,T_{1}\;\cos k_{x}\;\\
\xi_{\vec{k}}^{y} & =-2\lambda-\mu-2\,T_{1}\;\cos k_{y}\;\\
\gamma_{\vec{k}} & =4e^{i(k_{y}-k_{x})/2}\;\left(T_{2}\;\cos\frac{k_{x}}{2}\cos\frac{k_{y}}{2}\right.\\
 & \left.+T_{3}\;\cos\frac{3k_{x}}{2}\cos\frac{k_{y}}{2}+T_{3}\;\cos\frac{k_{x}}{2}\cos\frac{3k_{y}}{2}\right).
\end{align*}
The eigenvalues are given by $E_{\pm,\vec{k}}=\xi_{\vec{k}}\pm\sqrt{\eta_{\vec{k}}^{2}+|\gamma_{\vec{k}}|^{2}}$,
where $\xi_{\vec{k}}=(\xi_{\vec{k}}^{x}+\xi_{\vec{k}}^{y})/2$ and
$\eta_{\vec{k}}=(\xi_{\vec{k}}^{x}-\xi_{\vec{k}}^{y})/2$. For hole
doping $p$ (the number of fermions in our model) the lower band $E_{-,\vec{k}}$
will be partially occupied. The total area enclosed by the Fermi surface
in the lower band is equal to the hole doping $p$ (multiplied by
$4\pi^{2}$).

\begin{figure}
\begin{centering}
\includegraphics[scale=0.13]{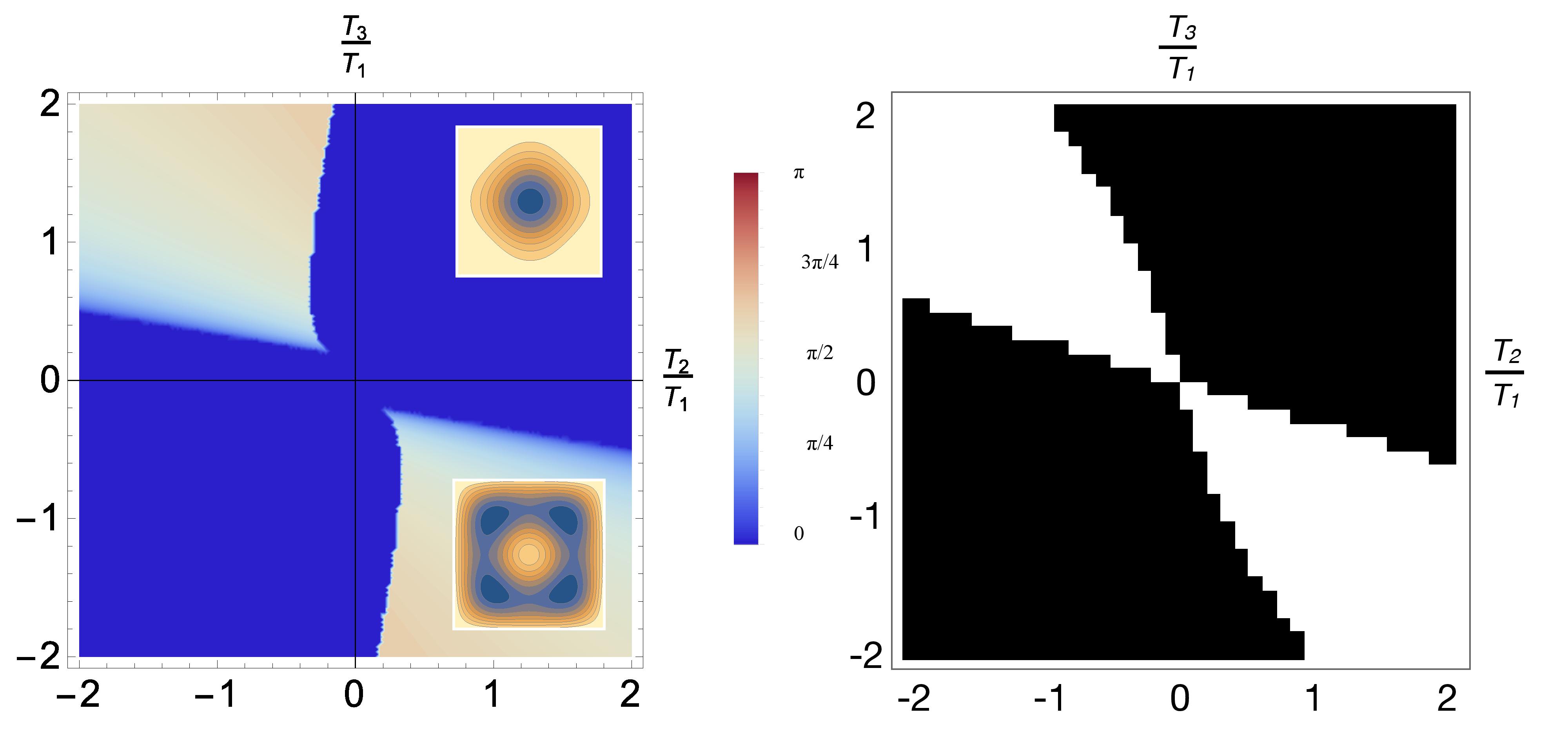} 
\par\end{centering}

\protect\protect\protect\caption{ \label{fig:k-minima} (A) Location of the band minima as a function
of $T_{3}/T_{1}$ and $T_{2}/T_{1}$. The color scale corresponds
to the distance along the $\Gamma-M$ line in the Brillouin zone:
blue corresponds to the $\Gamma$ point, $k_{x}=k_{y}=0$, and red
corresponds to the $M$ point, $k_{x}=k_{y}=\pi$. The insets show
contours of the dispersion of the lower band of the Hamiltonian Eq.~\eqref{eq:Hamiltonian_boson_fermion_meanfield}
for specific choices of parameters in the corresponding regions. (B)
Dominant superconductivity instability as a function of $T_{3}/T_{1}$
and $T_{2}/T_{1}$ for doping $p=0.25$ and $J=50$: d-wave (white)
\textit{vs.} s-wave (black). Note the good correlation between d-wave
superconductivity and the appearance of four band minima.}
\end{figure}

The Hamiltonian Eq.~\eqref{eq:H_k} has four-fold rotational symmetry,
$k_{x}\rightarrow k_{y}$ and $k_{y}\rightarrow-k_{x}$, and reflection
symmetry about the diagonals $k_{x}\rightarrow k_{y}$ and $k_{y}\rightarrow k_{x}$
as well as $k_{x}\rightarrow-k_{y}$ and $k_{y}\rightarrow-k_{x}$.
Depending on the relative values of $T_{1}$, $T_{2}$ and $T_{3}$,
the band minima will be located at different points in the Brillouin
zone, and the Fermi surface topology will vary accordingly. In Fig.~\ref{fig:k-minima}A
we show the position of the minima along the $k_{x}=\pm k_{y}$ directions
(or $\Gamma-M$ line), as a function of the ratios $T_{3}/T_{1}$
and $T_{2}/T_{1}$. We identify two regions in parameter space, where
the dispersion minima are (i) at the $\Gamma$ point (blue-colored
region), and (ii) in between the $\Gamma$ and $M$ points, varying
continuously with $T_{1}/T_{3}-T_{2}/T_{3}$ (faded region). An example
of dispersion where the minima are at $(k_{x},k_{y})\simeq(\pm\pi/2,\pm\pi/2)$
is shown in the bottom inset of Fig.~\ref{fig:k-minima}A. Case (ii)
is clearly conducive to the appearance of four Fermi pockets in an
appropriate range of the chemical potential. 

\textit{d-wave Superconductivity}. To study superconducting instabilities
we need to include four-fermion terms in the Hamiltonian, i.e., go
beyond the model introduced in Refs.~\onlinecite{key-30,key-32,key-34}
and summarized in Fig.~\ref{fig:configurations}B and Eq.~\eqref{eq:Hamiltonian_Total}.
Consider the $t-J$ Hamiltonian on the square lattice~\cite{key-36},
\begin{equation}
H_{tJ}=-\sum_{\alpha}t_{ij}d_{i,\alpha}^{\dagger}d_{j,\alpha}^{\;}+J\sum_{\left\langle i,j\right\rangle }\left(\vec{S}_{i}\cdot\vec{S}_{j}-\frac{1}{4}n_{i}n_{j}\right)\label{eq:t-J_model}
\end{equation}
subject to the constraint that $n_{i}\leq1$. Here $d_{i,\alpha}^{\dagger}$
and $d_{i,\alpha}^{\ }$ are the electron creation and annihilation
operators ($\alpha=\uparrow,\downarrow$) of the $t-J$ model, $\vec{S}_{i}=d_{i,\alpha}^{\dagger}\,\vec{\sigma}_{\alpha,\beta}\,d_{i,\beta}^{\ }$
(with $\alpha,\,\beta$ summed over), and $n_{i}=d_{i,\uparrow}^{\dagger}d_{i,\uparrow}^{\ }+d_{i,\downarrow}^{\dagger}d_{i,\downarrow}^{\ }$.

We can identify the dimer Hilbert space with a subspace of the Hilbert
space for the $t-J$ model, where the zero dimers state corresponds
to the state with zero electrons, and the rest of the Hilbert space
can be introduced via the operators $b_{i,\eta}^{\dagger}\Leftrightarrow\Upsilon_{i,\eta}\,(d_{i\uparrow}^{\dagger}d_{i+\eta\downarrow}^{\dagger}-d_{i\downarrow}^{\dagger}d_{i+\eta\uparrow}^{\dagger})/\sqrt{2}$
and $c_{i,\eta,\sigma}^{\dagger}\Leftrightarrow\Upsilon_{i,\eta}(d_{i,\sigma}^{\dagger}+d_{i+\eta,\sigma}^{\dagger})/\sqrt{2}$.
The phases $\Upsilon_{i,\eta}$ represent a gauge choice and we shall
follow the one by Rokhsar and Kivelson~\cite{key-1} and define $\Upsilon_{i,\hat{y}}=1$
and $\Upsilon_{i,\hat{x}}=\left(-1\right)^{i_{y}}$, where $i_{y}$
is the $y$-component of the 2D square lattice site index $i$.

Given the conventional inner product for the electron Hilbert space,
the dimer basis is not orthonormal. This issue can be addressed in
general by Gram-Schmidt orthogonalization~\cite{key-35}; however,
it is customary to use the leading order approximation and to assume
that the dimer states are orthogonal (and normalized)~\cite{key-33}.
The relevant Hamiltonian can then be determined by projecting Eq.~\eqref{eq:t-J_model}
onto this basis. The pairing term (four-fermion interaction) comes
from the spin-spin term in the $t-J$ model, namely $H_{J}=J\sum_{\langle i,j\rangle}\left(\vec{S}_{i}\cdot\vec{S}_{j}-\frac{1}{4}n_{i}n_{j}\right)$.
Let us focus on a single plaquette term and consider eight relevant
states for this plaquette, $c_{i,\hat{x},\alpha}^{\dagger}c_{i+\hat{y},\hat{x},\beta}^{\dagger}\left|0\right\rangle $
and $c_{i,\hat{y},\alpha}^{\dagger}c_{i+\hat{x},\hat{y},\beta}^{\dagger}\left|0\right\rangle $,
$\alpha,\,\beta=\uparrow,\downarrow$. The Hamiltonian $H_{J}$ is
non-zero only in the singlet channel and therefore we must restrict
the spins $\alpha,\beta$ to be in a singlet state, thereby the effective
Hamiltonian for the dimers is given by $\tilde{H}_{J}=\left(\begin{smallmatrix}-J/2 & 0\\
0 & -J/2
\end{smallmatrix}\right)$~\cite{key-33}. As such we add to our Hamiltonian in Eq.~\eqref{eq:Hamiltonian_Total}
the term: 
\begin{align}
\tilde{H}_{J} & =-\frac{J}{4}\sum_{i}\left(\epsilon_{\alpha\gamma}\,c_{i,\hat{x},\alpha}^{\dagger}\,c_{i+\hat{y},\hat{x},\gamma}^{\dagger}\right)\left(\epsilon_{\beta\delta}\,c_{i+\hat{y},\hat{x},\beta}^{\phantom{\dagger}}\,c_{i,\hat{x},\delta}^{\phantom{\dagger}}\right)\nonumber \\
 & +x\leftrightarrow y\,.\label{eq:Pairing_hamiltonian}
\end{align}

For convenience we define $\Delta_{x}=\epsilon_{\alpha\gamma}\langle c_{i,\hat{y},\alpha}^{\dagger}\,c_{i+\hat{x},\hat{y},\gamma}^{\dagger}\rangle$
and $\Delta_{y}=\epsilon_{\alpha\gamma}\langle c_{i,\hat{x},\alpha}^{\dagger}\,c_{i+\hat{y},\hat{x},\gamma}^{\dagger}\rangle$,
whereby d-wave pairing corresponds to $\Delta_{x}=-\Delta_{y}=\Delta$
(which in turn can be chosen real with an appropriate choice of phase).
Using a HS transformation on Eq.~\eqref{eq:Pairing_hamiltonian},
$H=\sum_{\vec{k}}\Psi_{\vec{k}}^{\dagger}\;{\cal H}_{\vec{k}}\;\Psi_{\vec{k}}^{\;}$,
where 
\begin{equation}
{\cal H}_{\vec{k}}=\begin{pmatrix}\xi_{\vec{k}}^{x} & \gamma_{\vec{k}} & \Delta_{\vec{k}}^{x} & 0\\
\gamma_{\vec{k}}^{*} & \xi_{\vec{k}}^{y} & 0 & \Delta_{\vec{k}}^{y}\\
\Delta_{\vec{k}}^{x} & 0 & -\xi_{-\vec{k}}^{x} & -\gamma_{-\vec{k}}\\
0 & \Delta_{\vec{k}}^{y} & -\gamma_{-\vec{k}}^{*} & -\xi_{-\vec{k}}^{y}\;,
\end{pmatrix}\label{eq:bogoluibov_de_gennes}
\end{equation}
and $\Psi_{\vec{k}}^{\dagger}=(c_{\vec{k},\hat{y},\uparrow}^{\dagger},c_{\vec{k},\hat{x},\uparrow}^{\dagger},c_{-\vec{k},\hat{y},\downarrow}^{\;},c_{-\vec{k},\hat{x},\downarrow}^{\;})$.
Here $\Delta_{\vec{k}}^{x}=\frac{J}{2}\Delta_{x}\cos k_{x}$ and $\Delta_{\vec{k}}^{y}=\frac{J}{2}\Delta_{y}\cos k_{y}$.
The eigenvalues of this Hamiltonian are given by 
\begin{equation}
E_{\pm,\pm,{\vec{k}}}=\pm\sqrt{\Theta_{\vec{k}}\pm\sqrt{\Lambda_{\vec{k}}+\Xi_{\vec{k}}}}\label{eq:Energy}
\end{equation}
where 
\begin{align*}
\Theta_{\vec{k}} & =\frac{1}{2}\left[(\xi_{\vec{k}}^{x})^{2}+(\xi_{\vec{k}}^{y})^{2}+(\Delta_{\vec{k}}^{x})^{2}+(\Delta_{\vec{k}}^{y})^{2}+2|\gamma_{\vec{k}}|^{2}\right]\\
\Lambda_{\vec{k}} & =\frac{1}{4}\left[(\xi_{\vec{k}}^{x})^{2}-(\xi_{\vec{k}}^{y})^{2}+(\Delta_{\vec{k}}^{x})^{2}-(\Delta_{\vec{k}}^{y})^{2}\right]^{2}\\
\Xi_{\vec{k}} & =|\gamma_{\vec{k}}|^{2}\left[(\xi_{\vec{k}}^{x}+\xi_{\vec{k}}^{y})^{2}+(\Delta_{\vec{k}}^{x}-\Delta_{\vec{k}}^{y})^{2}\right].
\end{align*}

When $T_{1}$, $T_{2}$ and $T_{3}$ are such that there are four
Fermi pockets (in the absence of superconductivity), there are eight
Dirac points in the dispersion, i.e., there are eight nodes where
the gap $E_{+,-,{\vec{k}}}=E_{-,-,{\vec{k}}}=0$. These points are
located along the diagonals of the Brillouin zone. When $k_{x}=\pm k_{y}$,
$\Lambda_{\vec{k}}$ vanishes, and the gap closing condition $\Theta_{\vec{k}}=\sqrt{\Xi_{\vec{k}}}$
is equivalent to $\xi_{{\vec{k}}}^{2}+\Delta_{{\vec{k}}}^{2}-|\gamma_{\vec{k}}|^{2}=0$,
where $\xi_{\vec{k}}=(\xi_{\vec{k}}^{x}+\xi_{\vec{k}}^{y})/2$ and
$\Delta_{\vec{k}}=(\Delta_{\vec{k}}^{x}-\Delta_{\vec{k}}^{y})/2$.
Notice that the Fermi surface in the absence of superconductivity
is given by $\xi_{{\vec{k}}}^{2}-|\gamma_{\vec{k}}|^{2}=0$. Therefore,
whenever there are four Fermi pockets, for a range of $\Delta_{{\vec{k}}}^{2}$
there will be two nodes for each pocket, slightly shifted along the
diagonal from the original Fermi surface~\cite{key-33}.

Using self consistent equations for the superconducting order parameter,
we can then compare s-wave and d-wave instabilities. Up to an unimportant
constant energy shift, the Gibbs free energy is given by 
\begin{align}
G & =\frac{J}{4}\left(\left|\Delta_{x}^{2}\right|+\left|\Delta_{y}^{2}\right|\right)\label{eq:Gibbs_free_energy}\\
 & -\frac{2}{\beta}\int\frac{d^{2}k}{4\pi^{2}}\ln\left[\cosh\left(\frac{\beta}{2}E_{+,+,\vec{k}}\right)\cosh\left(\frac{\beta}{2}E_{+,-,\vec{k}}\right)\right].\nonumber 
\end{align}
Minimizing the free energy with respect to $\Delta_{x}$, we obtain:
\begin{align}
\Delta_{x} & \!=\!\!\sum_{s=\pm}\int\frac{d^{2}k}{4\pi^{2}}\;\frac{\tanh\left(\frac{\beta}{2}E_{+,s,\vec{k}}\right)\cos\left(k_{x}\right)}{2{E}_{+,s,\vec{k}}}\label{eq:Self_consistency}\\
 & \!\!\times\left\{ \Delta_{\vec{k}}^{x}+\frac{s}{\sqrt{\Lambda_{\vec{k}}+\Xi_{\vec{k}}}}\left[\sqrt{\Lambda_{\vec{k}}}\;\Delta_{\vec{k}}^{x}+\left|\gamma_{\vec{k}}\right|^{2}\left(\Delta_{\vec{k}}^{x}-\Delta_{\vec{k}}^{y}\right)\right]\right\} \nonumber 
\end{align}
and similarly for $\Delta_{y}$. From the symmetries of this equation
we see that there are two solutions, $\Delta_{x}=\mp\Delta_{y}$,
corresponding to d-wave and extended s-wave superconductivity.

We numerically compare the two solutions at zero temperature and find
that d-wave superconductivity wins for a large range of ratios $T_{2}/T_{1}$
and $T_{3}/T_{1}$, as illustrated in Fig.~\ref{fig:k-minima}B.
The correlation between the region with fermion pockets depicted in
Fig~\ref{fig:k-minima}A and the region with d-wave superconductivity
in Fig.~\ref{fig:k-minima}B is evident. This can be qualitatively
understood as the largest change in the Gibbs free energy upon entering
the superconducting state comes from the contribution of the integral
around the FS. Since the shape of the four Fermi pockets follows largely
the nodal lines of the s-wave order parameter, and it anti-correlates
with the d-wave nodal lines, one expects the appearance of the pockets
to favor d-wave superconductivity.

Whereas the horizontal boundaries match very well in the two panels
in Fig.~\ref{fig:k-minima}, the vertical boundaries less so. Indeed,
along the horizontal boundary the dispersion transitions smoothly
from having a single minimum at the $\Gamma$ point to having four
minima along the $\Gamma-M$ direction in the Brillouin zone, i.e.,
the minima move continuously away from the $\Gamma$ point (which
thus becomes a maximum). On the other hand, along the vertical boundary,
the minima jump discontinuously from the $\Gamma$ point to the new
four minima, as four local minima at finite momenta dip down to become
the global minima. Depending on the value of the chemical potential,
there is a region in the $T_{2}/T_{1}$ \textit{vs.} $T_{3}/T_{1}$
plane near the vertical boundary where the Fermi surface has five
sheets, four pockets coexisting with a surface surrounding the $\Gamma$
point. The latter favors s-wave superconductivity as it has no nodes
at the $\Gamma$ point, and it is therefore expected to shift the
position of the boundary between d-wave and s-wave superconductivity,
as observed.

\textit{Conclusions.} We presented a slave particle formulation of
a mixed boson+fermion quantum dimer model recently proposed in the
context of high-T$_{c}$ superconductors~\cite{key-30,key-32,key-34}.
A key finding of this work is that substantial progress can be made
using a mean-field analysis that simply assumes a translational and
rotational invariant (liquid) state for the bosonic component. We
analyze the effective theory for the remaining fermionic degrees of
freedom, and distinguish between two regimes of Fermi surface topology,
depending on the effective couplings obtained from both microscopic
parameters and correlations of the bosonic liquid state. The two regimes
correspond to one Fermi surface around the $\Gamma$ point, or four
Fermi pockets centered along the $\Gamma-M$ lines. By including additional
interactions that arise from the $t-J$ model, we find that the system
is unstable to superconductivity. The symmetry of the superconducting
order parameter, s-wave \textit{vs.} d-wave, is shown to correlate
strongly with the Fermi surface topology, with d-wave being favored
when four Fermi pockets are present.

\textbf{Acknowledgements:} This work was supported in part by Engineering
and Physical Sciences Research Council (EPSRC) Grants No. EP/G049394/1
(C.Ca.) and No. EP/M007065/1 (C.Ca. and G.G.), by DOE Grant DEF-06ER46316
(C.Ch.), and by the EPSRC Network Plus on ``Emergence and Physics
far from Equilibrium''. Statement of compliance with the EPSRC policy
framework on research data: this publication reports theoretical work
that does not require supporting research data. C.Ca. and G.G. thank
the BU visitor program for its hospitality.

\part*{\newpage{}Supplementary Online Information}

\section{\label{sec:Effective-Hamiltonian}Effective Hamiltonian }

Here we provide more details about the derivation of the effective
two body (four fermion) interaction introduced in Eq.~\eqref{eq:Pairing_hamiltonian}
in the main text. The procedure to obtain this term for the dimer
model is to identify the dimer Hilbert space with a subspace of the
$t-J$ model Hilbert space and project the $t-J$ Hamiltonian Eq.~\eqref{eq:t-J_model}
accordingly.

We identify the state with zero dimers of any kind with the state
with zero electrons for the $t-J$ model. The rest of the Hilbert
space for the dimers can be introduced via the operators $b_{i,\eta}^{\dagger}\Leftrightarrow\Upsilon_{i,\eta}\,(d_{i\uparrow}^{\dagger}d_{i+\eta\downarrow}^{\dagger}-d_{i\downarrow}^{\dagger}d_{i+\eta\uparrow}^{\dagger})/\sqrt{2}$
and $c_{i,\eta,\sigma}^{\dagger}\Leftrightarrow\Upsilon_{i,\eta}(d_{i,\sigma}^{\dagger}+d_{i+\eta,\sigma}^{\dagger})/\sqrt{2}$.
The phases $\Upsilon_{i,\eta}$ represent a gauge choice and we shall
follow the one by Rokhsar and Kivelson \cite{key-1-1} and define
$\Upsilon_{i,\hat{y}}=1$ and $\Upsilon_{i,\hat{x}}=\left(-1\right)^{i_{y}}$,
here $i_{y}$ is the y-component of the dimer co-ordinate. Given the
conventional inner product for the electron Hilbert space, the dimer
basis is not orthonormal and therefore does not serve as a convenient
basis to calculate matrix elements. This can be resolved by Gram-Schmidt
orthogonalization. In general, if we denote the basis elements of
the dimer Hilbert space by $|A\rangle$ and the overlap matrix between
states $S_{AB}=\langle A|B\rangle$, then an orthonormal basis for
the Hilbert space is given by~\cite{key-35-1}: 
\begin{equation}
\vert\tilde{A}\rangle=\sum_{A}\left(S^{-1/2}\right)_{A,\tilde{A}}|A\rangle.\label{eq:Basis}
\end{equation}
It is not too hard to check that the matrix $S$ is a real symmetric
matrix $S_{AB}=S_{BA}$ and therefore $S^{\dagger}=S$. From this
it follows that $\left\langle \tilde{A}\mid\tilde{B}\right\rangle =S^{-1/2}SS^{-1/2}=\delta_{\tilde{A},\tilde{B}}$,
i.e., the new states are orthonormal.

The Hamiltonian projected onto this basis is given by~\cite{key-35-1}:
\begin{equation}
H_{\tilde{A},\tilde{B}}=\sum_{A,B}(S^{-1/2})_{\tilde{A},A}\;\langle A|H_{tJ}|B\rangle\;(S^{-1/2})_{B,\tilde{B}}.\label{eq:Hamiltonian}
\end{equation}
To leading order, $S_{AB}\simeq\delta_{A,B}$ as the dimers are nearly
orthogonal. To show this consider two states $\left|A\right\rangle $
and $\left|B\right\rangle $. We can form the loop graph of $\left|A\right\rangle $
and $\left|B\right\rangle $ by deleting all the dimers that $\left|A\right\rangle $
and $\left|B\right\rangle $ have in common. The rest of the dimers
will form loops (with dimers from state $\left|A\right\rangle $ and
state $\left|B\right\rangle $ alternating along a loop). If there
is a loop of length 2, that is two dimers of different type on the
same link then $\left\langle A\mid B\right\rangle =0$, so we have
$S_{AB}=\delta_{A,B}$ for those states. Assuming there are no such
links we have that all loops are at least length four. Now the overlap
of $\left|A\right\rangle $ and $\left|B\right\rangle $ is the product
of overlaps over all loops. Furthermore it is known that the overlap
between two loops is exponential in the length of the loop \cite{key-1-1,key-35-1}.
Since all loops are of at least length four (rather long) to leading
order we may set the overlap matrix to zero if there is at least one
loop or the states $\left|A\right\rangle $ and $\left|B\right\rangle $
are different. Now the states are normalized to unity so we have $S_{AB}\simeq\delta_{A,B}$.

The pairing term (four-fermion interaction) comes from the spin-spin
term in the $t-J$ model, namely $H_{J}=J\sum_{\langle i,j\rangle}\left(\vec{S}_{i}\cdot\vec{S}_{j}-\frac{1}{4}n_{i}n_{j}\right)$.
Let us focus on a single plaquette term and consider eight relevant
states for the dimers on this plaquette, $c_{i,\hat{x},\alpha}^{\dagger}c_{i+\hat{y},\hat{x},\beta}^{\dagger}\left|0\right\rangle $
and $c_{i,\hat{y},\alpha}^{\dagger}c_{i+\hat{x},\hat{y},\beta}^{\dagger}\left|0\right\rangle $.
We notice that the Hamiltonian $H_{J}$ is zero in the triplet channel.
This means that the effective Hamiltonian for the dimers $\tilde{H}_{J}$
is also zero in the triplet channel. Indeed, the spins of the dimers
are the same as the spins of the electrons for the $t-J$ model, so
the projected Hamiltonian has the same spin structure. As such we
might as well restrict the spins of the dimers to lie in a singlet
(there are two such states per plaquette with the two dimers lying
either along the x-axis or along the y-axis). Moreover, the projected
Hamiltonian is diagonal in this basis. Indeed the bare Hamiltonian
$H_{J}$ contains no hopping terms for the electrons, only spin flip
terms. As such the only terms that could contribute to off diagonal
matrix elements come from states of the form $d_{i,\alpha}^{\dagger}d_{i+\hat{x}+\hat{y},\beta}^{\dagger}\left|0\right\rangle $
and $d_{i+\hat{x},\alpha}^{\dagger}d_{i+\hat{y},\beta}^{\dagger}\left|0\right\rangle $
(and linear combinations thereof) which belong to both dimer configurations
(along the x-axis and along the y-axis). However the Hamiltonian annihilates
such states and the projected Hamiltonian has no corresponding hopping
terms. By symmetry the Hamiltonian when restricted to the singlet
subspace is a multiple of the identity matrix. Its value is given
by: \begin{widetext} 
\begin{align*}
 & 2\left\langle 0\right|\frac{1}{2\sqrt{2}}\left\{ \left(d_{i,\uparrow}+d_{i+\hat{x},\uparrow}\right)\left(d_{i+\hat{y},\downarrow}+d_{i+\hat{y}+\hat{x},\downarrow}\right)-\left(d_{i,\downarrow}+d_{i+\hat{x},\downarrow}\right)\left(d_{i+\hat{y},\uparrow}+d_{i+\hat{y}+\hat{x},\uparrow}\right)\right\} J\left(\vec{S}_{i}\cdot\vec{S}_{i+\hat{y}}-\frac{1}{4}n_{i}n_{i+\hat{y}}\right)\\
 & \left\{ \left(d_{i+\hat{y},\downarrow}^{\dagger}+d_{i+\hat{y}+\hat{x},\downarrow}^{\dagger}\right)\left(d_{i,\uparrow}^{\dagger}+d_{i+\hat{x},\uparrow}^{\dagger}\right)-\left(d_{i+\hat{y},\uparrow}^{\dagger}+d_{i+\hat{y}+\hat{x},\uparrow}^{\dagger}\right)\left(d_{i,\downarrow}^{\dagger}+d_{i+\hat{x},\downarrow}^{\dagger}\right)\right\} \frac{1}{2\sqrt{2}}\left|0\right\rangle =-\frac{J}{2}
\end{align*}
\end{widetext} and, within the spin-singlet channel, the Hamiltonian
is $\tilde{H}_{J}=\left(\begin{smallmatrix}-J/2 & 0\\
0 & -J/2
\end{smallmatrix}\right)$. Correspondingly, we can add to our Hamiltonian in Eq.~\eqref{eq:Hamiltonian_Total}
the term: 
\begin{align}
\tilde{H}_{J} & =
-\frac{J}{4}\sum_{i}\left(\epsilon_{\alpha\gamma}\;c_{i,\hat{x},\alpha}^{\dagger}\,c_{i+\hat{y},\hat{x},\gamma}^{\dagger}\right)\left(\epsilon_{\beta\delta}c_{i+\hat{y},\hat{x},\beta}^{\phantom{\dagger}}\,c_{i,\hat{x},\delta}^{\phantom{\dagger}}\right)\nonumber \\
 & +x\leftrightarrow y\,.\label{eq:Pairing_hamiltonain}
\end{align}
This is a spin spin Hamiltonian for the fermionic dimers. 

\section{\label{sec:Dirac-Cones}Dirac Cones}

\begin{figure}
\begin{centering}
\includegraphics[scale=0.22]{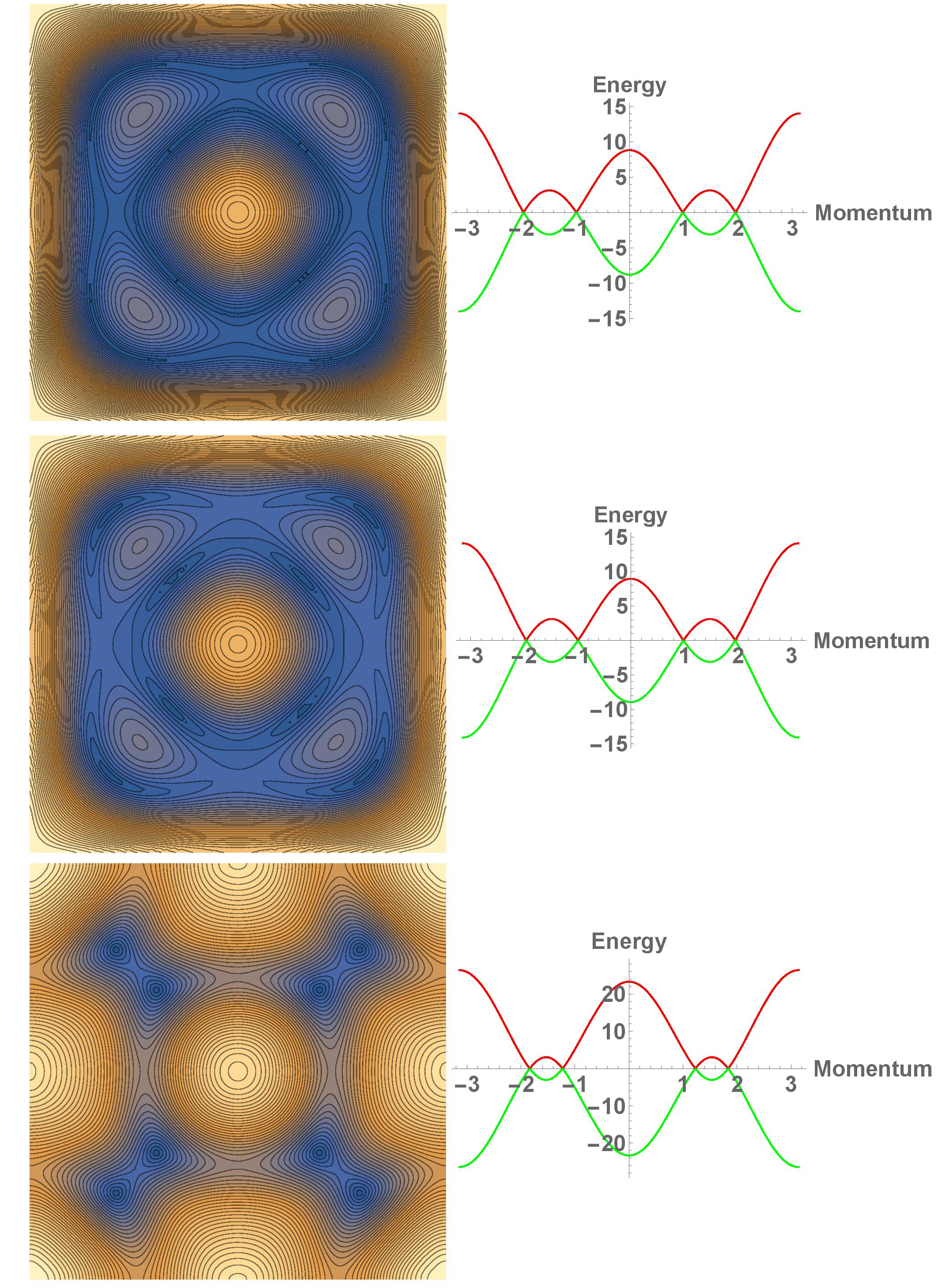} 
\par\end{centering}

\protect\protect\protect\caption{\label{fig:Dirac_cones} Particular realization of the Dirac Cones
for $T_{1}=1$, $T_{2}=-3.9$, $T_{3}=1.8$, and $J/T_{1}=18$ (top),
$27$ (middle), and $90$ (bottom panel). The superconducting paring
was chosen to optimize the zero temperature Gibbs free energy $\Delta\sim0.05,\,0.135,\,0.5$,
respectively. The whole Brillouin zone from $(-\pi,-\pi)$ to $(\pi,\pi)$
is shown on the left and a cross section along the major diagonal
on the right, which highlights $4$ of the Dirac cones.}
\end{figure}

To verify the existence and robustness of the Dirac cones predicted
in the main text, we plot numerically the function $E_{+,-}\left(k\right)$
near the values of the hopping matrix elements $T_{1}=1$, $T_{2}=-3.9$,
$T_{3}=1.8$, for various values of $J$ near the $\Delta$ that minimizes
the free energy. We focus on the case of large $J/T_{1}=18,\,27,\,90$
(see Fig.~\ref{fig:Dirac_cones}), and we find eight cones in all
cases, even though the exchange coupling is taken to be $\sim5$ times
larger then the relevant value for cuprate superconductors.

To get an estimate of the ratio of $J/T_{1}$ we note that for the
cuprates $J\sim0.13$~eV, the nearest neighbor hopping $t\sim0.4$~eV~\cite{key-36-1,key-40-1,key-48-1},
$t_{1}\simeq0.35t$~\cite{key-30-2}, and in general $\left\langle b_{i_{1},\eta_{1}}^{\dagger}b_{i_{2},\eta_{2}}\right\rangle <\frac{1}{4}\left(1-p\right)$,
as any boson bilinear must be less then the largest occupation number
$\left\langle n_{i}\right\rangle =\frac{1}{4}\left(1-p\right)$. Recalling
that $T_{1}=t_{1}\left\langle b_{i,\hat{x}}^{\dagger}b_{i+\hat{y},\hat{x}}\right\rangle $,
we get for $p\simeq0.2$ that $J/T_{1}\gtrsim5$, likely of the order
of $10-20$.

\section{\label{sec:Large-N-arbitrary}Large N arbitrary S}

The considerations given in the main text can be extended to multiple
species ($N$) and different occupation number ($S$) of dimers. The
meanfield studied in the main text becomes arbitrarily accurate in
this limit. We consider the case of the square lattice but generalizations
to different lattice geometries is straightforward. We consider $N$
different species of bosonic dimers $\gamma=1,2,...N$ living on the
links of the lattice and $2N$ species of fermionic dimers $\sigma,\gamma=\uparrow1;\uparrow2;\ldots\uparrow N;\downarrow1\;\ldots\downarrow N$
also living on the same links. We represent the dimers using $N$
species of bosons living on the links of the lattice $b_{i,\eta,\gamma}$
and $2N$ species of fermions $c_{i,\eta,\sigma,\gamma}$. The dimer
Hilbert space can be mapped onto a subspace of the boson/fermion Hilbert
space via the correspondence where $n_{i,\eta,\gamma}$ bosonic dimers
of species $\gamma$ living on the link $i,i+\eta$ are identified
with the state for the slave particles where the occupation number
of the boson $\gamma$ on link $i,i+\eta$ is given by $n_{i,\eta,\gamma}$.
Similarly for the fermions. Rather than enforcing the constraint of
at most one dimer per link (which is redundant in the physical case
where $N=S=1$), we introduce the constraint 
\begin{equation}
\Pi_{i}^{N,S}\equiv\sum_{\gamma=1}^{N}\sum_{l\in v_{i}}n_{l,\gamma}+n_{l,\uparrow,\gamma}+n_{l,\downarrow,\gamma}-NS=0\,.\label{eq:Constraint-1}
\end{equation}
Here, for convenience of notation, $\ell\in v_{i}$ labels the four
links $j,\eta$ that are attached to vertex $i$. In the dimer language
this corresponds to the constraint that the total number of dimers
of any species on all the links touching the vertex $i$ is given
by $NS$. $S$ can be an arbitrary number and in the limit where $S$
becomes large the bosonic part of the dimer model becomes semiclassical.
In the path integral formulation, the constraint can be written as
\begin{widetext} 
\begin{equation}
\Pi_{i}^{N,S}\sim\int d\lambda_{i}\exp\left[-\varepsilon\lambda_{i}\left(\sum_{\gamma=1}^{N}\sum_{l\in v_{i}}n_{l,\gamma}+n_{l,\uparrow,\gamma}+n_{l,\downarrow,\gamma}-NS\right)\right]\,,\label{eq:Constarint_integral}
\end{equation}
\end{widetext} where $\varepsilon$ is the length of a time slice.

There is no prescription to write a Hamiltonian for a large $N$ theory.
However, in order to proceed further we need to write down Hamiltonians
for our slave bosons/fermions that reduce to the Hamiltonian given
in the main text in the case $N,S=1$ and are amenable to large $N$
expansion~\cite{key-9}. All the Hamiltonians written in this section
have a direct interpretation in terms of Hamiltonians for the dimers
(they all correspond to various dimer hopping terms and terms that
count the number of flippable dimer plaquettes). Note that it does
not matter whether the Hamiltonians we produce respect the constraint
in Eq.~\eqref{eq:Constraint-1} as in the path integral formulation
we insert projectors onto the physical space $\Pi_{i}^{N,S}$ at every
time slice. The hint towards how to do this extension comes from the
expressions derived around Eqs.~\eqref{eq:Hubbard_Stratonovich_potential}
and~\eqref{eq:Hubbard_Stratonovich_kinetic} in the main text. Indeed
in order to apply a HS transformation to our expressions we need to
write our Hamiltonian (when restricted to a single plaquette) schematically
in the form $H_{D}=AB$ where $A$ and $B$ are single particle operators
for one species of dimer, either bosonic or fermionic (see the main
text e.g. Eqs.~\eqref{eq:Hubbard_Stratonovich_potential}, \eqref{eq:Hubbard_Stratonovich_kinetic}
and~\eqref{eq:Hamiltonian_boson_fermion_meanfield}). The main idea
is to replace $H_{D}\rightarrow H_{D}^{NS}=\frac{1}{N}\sum_{\gamma_{1}=1}^{N}A_{\gamma_{1}}\sum_{\gamma_{2}=1}^{N}B_{\gamma_{2}}$.
Here $A_{\gamma_{1}}$, $B_{\gamma_{2}}$ are single particle operators
either bosonic or fermionic which are identical to $A$ and $B$ except
they now carry an index $\gamma$. As such each HS transformation
given in the main text corresponds to a different large $N$ Hamiltonian.
In the large $N$ limit, with this extension, we present models where
the HS mean-field becomes arbitrarily accurate. Qualitatively we expect
mean-field theory to become more and more accurate as each particle
interacts with $N$ particles with an interaction strength that is
attenuated by $1/N$. We now proceed to give several examples of this
procedure. We note that none of the Hamiltonians have any dependence
on $S$. In particular we can replace \begin{widetext} 
\begin{align}
 & Vb_{i,\hat{x}}^{\dagger}b_{i,\hat{x}}b_{i+\hat{y},\hat{x}}^{\dagger}b_{i+\hat{y},\hat{x}}\rightarrow\nonumber \\
 & \rightarrow\kappa_{1}\frac{V}{N}\left\{ \sum_{\gamma_{1}}b_{i,\hat{x},\gamma_{1}}^{\dagger}b_{i,\hat{x},\gamma_{1}}\cdot\sum_{\gamma_{2}}b_{i+\hat{y},\hat{x},\gamma_{2}}^{\dagger}b_{i+\hat{y},\hat{x},\gamma_{2}}\right\} +\left(1-\kappa_{1}\right)\frac{V}{N}\left\{ \sum_{\gamma_{1}}b_{i,\hat{x},\gamma_{1}}^{\dagger}b_{i+\hat{y},\hat{x},\gamma_{1}}^{\dagger}\cdot\sum_{\gamma_{2}}b_{i,\hat{x},\gamma_{2}}b_{i+\hat{y},\hat{x},\gamma_{2}}\right\} \label{eq:Hubbard_Stratonovich_potential-1}
\end{align}
where $\kappa_{1}$ is arbitrary and each value of $\kappa_{1}$ produces
a different Hamiltonian. Similarly we have 
\begin{align}
 & Jb_{i,\hat{x}}^{\dagger}b_{i+\hat{y},\hat{x}}^{\dagger}b_{i,\hat{y}}b_{i+\hat{x},\hat{y}}+h.c\rightarrow\nonumber \\
 & \rightarrow\kappa_{2}\frac{J}{N}\left\{ \sum_{\gamma_{1}}b_{i,\hat{x},\gamma_{1}}^{\dagger}b_{i+\hat{y},\hat{x},\gamma_{1}}^{\dagger}\cdot\sum_{\gamma_{2}}b_{i,\hat{y},\gamma_{2}}b_{i+\hat{x},\hat{y},\gamma_{2}}+h.c.\right\} +\left(1-\kappa_{2}\right)\frac{J}{N}\left\{ \sum_{\gamma_{1}}b_{i,\hat{x},\gamma_{1}}^{\dagger}b_{i,\hat{y}}\cdot\sum_{\gamma_{2}}b_{i+\hat{y},\hat{x},\gamma_{2}}^{\dagger}b_{i+\hat{x},\hat{y},\gamma_{2}}+h.c.\right\} \label{eq:Hubbard_Stratonovich_kinetic-1}
\end{align}
where again $\kappa_{2}$ is arbitrary. The Bose/Fermi part may be
extended to the large $N$ limit in a similar manner. For instance
\begin{align}
 & \sum_{\sigma}\sum_{i}\left\{ -t_{1}b_{i,\hat{x}}^{\dagger}c_{i+\hat{y},\hat{x},\sigma}^{\dagger}c_{i,\hat{x},\sigma}b_{i+\hat{y},\hat{x}}+3\:terms\right.\nonumber \\
 & -t_{2}b_{i+\hat{x},\hat{y}}^{\dagger}c_{i,\hat{y},\sigma}^{\dagger}c_{i,\hat{x},\sigma}b_{i+\hat{y},\hat{x}}+7\:terms\nonumber \\
 & -t_{3}b_{i+\hat{x}+\hat{y},\hat{x}}^{\dagger}c_{i,\hat{y},\sigma}^{\dagger}c_{i+\hat{x}+\hat{y},\hat{x},\sigma}b_{i,\hat{y}}+7\:terms\nonumber \\
 & \left.-t_{3}b_{i+2\hat{y},\hat{x}}^{\dagger}c_{i,\hat{y},\sigma}^{\dagger}c_{i+2\hat{y},\hat{x},\sigma}b_{i,\hat{y}}+7\:terms\right\} \rightarrow\nonumber \\
 & \rightarrow\frac{1}{N}\sum_{\alpha}\sum_{i}\left\{ -t_{1}\sum_{\gamma_{1}}c_{i+\hat{y},\hat{x},\alpha,\gamma_{1}}^{\dagger}c_{i,\hat{x},\alpha,\gamma_{1}}\cdot\sum_{\gamma_{2}}b_{i,\hat{x},\gamma_{2}}^{\dagger}b_{i+\hat{y},\hat{x},\gamma_{2}}+3\;terms\right.\nonumber \\
 & -t_{2}\sum_{\gamma_{1}}c_{i,\hat{y},\alpha,\gamma_{1}}^{\dagger}c_{i,\hat{x},\alpha,\gamma_{1}}\cdot\sum_{\gamma_{2}}b_{i+\hat{x},\hat{y},\gamma_{2}}^{\dagger}b_{i+\hat{y},\hat{x},\gamma_{2}}+7\;terms\nonumber \\
 & -t_{3}\sum_{\gamma_{1}}c_{i,\hat{y},\alpha,\gamma_{1}}^{\dagger}c_{i+\hat{x}+\hat{y},\hat{x},\alpha,\gamma_{1}}\sum_{\gamma_{2}}b_{i+\hat{x}+\hat{y},\hat{x},\gamma_{2}}^{\dagger}b_{i,\hat{y},\gamma_{2}}+7\;terms\nonumber \\
 & \left.-t_{3}\sum_{\gamma_{1}}c_{i,\hat{y},\alpha,\gamma_{1}}^{\dagger}c_{i+2\hat{y},\hat{x},\alpha,\gamma_{1}}\sum_{\gamma_{2}}b_{i+2\hat{y},\hat{x},\gamma_{2}}^{\dagger}b_{i,\hat{y},\gamma_{2}}+7\;terms\right\} .\label{eq:Bose_Fermi}
\end{align}

The four fermion term in the Hamiltonian (see Eq.~\eqref{eq:Pairing_hamiltonian})
is given by spin spin interactions and has a variety of large $N$
extensions which were previously tabulated~\cite{key-11,key-12,key-13,key-10-1,key-9}.
We will not repeat this procedure here. We note that, to produce superconductivity,
the optimum extension is given by the $SP(2N)$ formalism~\cite{key-11,key-12,key-13}.
It is possible to do a HS decoupling on all these terms, based on
the following identity 
\begin{equation}
\exp\left(-\frac{\varepsilon}{N}\sum_{\gamma_{1}=1}^{N}A_{\gamma_{1}}\sum_{\gamma_{2}=1}^{N}B_{\gamma_{2}}\right)\sim\int dx_{1}\int dx_{2}\:\exp\left[\varepsilon N\left(x_{1}x_{2}-x_{1}\frac{1}{N}\sum_{\gamma_{2}}B_{\gamma_{2}}-x_{2}\frac{1}{N}\sum_{\gamma_{1}}A_{\gamma_{1}}\right)\right]\,.\label{eq:Hubbard_Stratonovich_Large_N}
\end{equation}
\end{widetext}

The mean-field equations (stationary points of this integral) are
given by $x_{2}=\frac{1}{N}\sum_{\gamma_{2}}\left\langle B_{\gamma_{2}}\right\rangle $
and $x_{1}=\frac{1}{N}\sum_{\gamma_{1}}\left\langle A_{\gamma_{1}}\right\rangle $.
From this we see that the mean-field is simply a sum of $N$ copies
of the mean-field for a single species problem and we can replace
$x_{2}=\left\langle B\right\rangle $ and $x_{1}=\left\langle A\right\rangle $,
where $A$ and $B$ are single particle operators. There is one saddle
point which reduces to $N$ copies of the mean-field theories found
in the main text.

In the limit where $N$ goes to infinity, the mean-field results become
exact, see Ref.~\onlinecite{key-9} (section 17.2). The derivation
given in Ref.~\onlinecite{key-9} works verbatim for our case.
Indeed an integration over the bosonic and fermionic fields that appear
in the partition function of our theory (for which the Hamiltonian
is quadratic) may be performed to obtain: 
\begin{equation}
Z=\int D\lambda_{i}D\left\{ x_{i,j}\right\} \:\exp\left[N\left(\lambda_{i},\left\{ x_{i,j}\right\} \right)\right]\,,\label{eq:Action_large_N}
\end{equation}
where $x_{i,j}$ are all the possible HS fields which we may introduce.
The only dependence on $N$ is the overall scaling $\sim N$ of the
action. Using an argument identical to Ref.~\onlinecite{key-9}
(section 17.2) it is possible to show that higher loop corrections
to the partition function in Eq.~\eqref{eq:Action_large_N} vanish
as $\left(1/N\right)^{P-L}$, where $P$ is the number of propagators
and $L$ is the number of loops. As such all the higher loop corrections
vanish when $N\rightarrow\infty$, making mean-field exact.

\section{\label{sec:Gauge-Symmetry-and}Gauge symmetry }

To discuss the various symmetries of our systems we focus for simplicity
on the case when $N=1$ for arbitrary $S$ (this does not entail any
additional complexity beyond the physical case of $S=1$). Consider
the gauge transformation where we assign a $U\left(1\right)$ phase
to each vertex of our system, namely where each boson and fermion
operator $b_{i,\eta}$ and $c_{i,\eta,\sigma}$ on each link transforms
as: 
\begin{equation}
b_{i,\eta}\rightarrow e^{i\theta_{i}}b_{i,\eta}e^{i\theta_{i+\eta}},\quad c_{i,\eta,\sigma}\rightarrow e^{i\theta_{i}}c_{i,\eta,\sigma}e^{i\theta_{i+\eta}}\,.\label{eq:Gauge_transformation-1}
\end{equation}
Any Hamiltonian that preserves the constraint 
\begin{equation}
\Pi_{i}^{S}\equiv\sum_{i,\eta\in i}n_{i,\eta}+n_{i,\eta,\uparrow}+n_{i,\eta,\downarrow}-S=0\label{eq:Constraint-2}
\end{equation}
is automatically invariant under the $U\left(1\right)$ Gauge transformation
in Eq.~\eqref{eq:Gauge_transformation-1}. Indeed any Hamiltonian
that preserves the constraint in Eq.~\eqref{eq:Constraint-2} can
be written as a sum of monomials each of which is a product of creation
and annihilation operators. In order to preserve the constraint Eq.~\eqref{eq:Constraint-2}
we mush have the same number of creation and annihilation operators
for the bosons/fermions at every vertex (otherwise the constraint
is no longer satisfied). Under this condition, the total phase from
phase factors in Eq.~\eqref{eq:Gauge_transformation-1} associated
with each vertex vanishes, leading to an invariant Hamiltonian. In
particular, one can explicitly check that the Hamiltonians in Eqs.~\eqref{eq:Hamiltonian_Total}
in the main text are invariant under the gauge transformation given
in Eq.~\eqref{eq:Gauge_transformation-1}.

This gauge transformation is compatible with many of the HS transformations
introduced in the main text. For instance the HS transformation associated
with the decoupling of the bosons, \begin{widetext} 
\begin{align}
H_{MF} & =V\left(\sum_{P}\left(x_{i2}n_{i+\hat{y},\hat{x}}+n_{i,\hat{x}}x_{i1}-x_{i1}x_{i2}\right)+\sum_{P}\left(x_{i3}n_{i,\hat{y}}+n_{i+\hat{x}}x_{i4}-x_{i3}x_{i4}\right)\right)\nonumber \\
 & +t\sum_{i}\kappa\left(z_{i1}^{*}b_{i,\hat{y}}b_{i+\hat{x},\hat{y}}+b_{i,\hat{x}}^{\dagger}b_{i+\hat{y},\hat{x}}^{\dagger}z_{i2}-z_{i1}^{*}z_{i2}+h.c.\right)+\frac{t}{2}\sum_{i}\left(1-\kappa\right)\left(w_{i1}^{*}b_{i+\hat{y},\hat{x}}^{\dagger}b_{i,\hat{y}}+\right.\nonumber \\
 & \left.+b_{i,\hat{x}}^{\dagger}b_{i+\hat{x},\hat{y}}w_{i2}-w_{i1}^{*}w_{i2}+h.c.\right)+\frac{t}{2}\sum_{i}\left(1-\kappa\right)\left(w_{i3}^{*}b_{i+\hat{y},\hat{x}}^{\dagger}b_{i+\hat{x},\hat{y}}+b_{i,\hat{x}}^{\dagger}b_{i,\hat{y}}w_{i4}-w_{i3}^{*}w_{i4}+h.c.\right)\,,\label{eq:Mean_field_gauage}
\end{align}
\end{widetext} is gauge compatible (here $\kappa$ is an arbitrary
constant). One simply has to transform the variables $z_{i,j}$ and
$w_{i,j}$ in the opposite way as the gauge transformation for the
bosons. Focusing on a single plaquette and labeling the lower left
corner site as $i$ with the other sites $i+\hat{x}$, $i+\hat{y}$
and $i+\hat{x}+\hat{y}$, the gauge transformation for the HS bosons
is given by: 
\begin{align}
x_{i,j} & \rightarrow x_{i,j}\nonumber \\
z_{i,j} & \rightarrow\exp\left[i\left(\theta_{i}+\theta_{i+\hat{x}}+\theta_{i+\hat{y}}+\theta_{i+\hat{x}+\hat{y}}\right)\right]z_{i,j}\nonumber \\
w_{i1/2} & \rightarrow\exp\left[i\left(\theta_{i}-\theta_{i+\hat{x}+\hat{y}}\right)\right]w_{i1/2}\nonumber \\
w_{i3/4} & \rightarrow\exp\left[i\left(\theta_{i+\hat{x}}-\theta_{i+\hat{y}}\right)\right]w_{i3/4}\,.\label{eq:Gauge_transformation_hubbard_Stratanovich}
\end{align}

Similar considerations can be made about the boson fermion and the
fermion fermion part of the Hamiltonian in the main text (in particular
the HS transformations in Eqs.~\eqref{eq:Hamiltonian_boson_fermion_meanfield}
and~\eqref{eq:Pairing_hamiltonian} are completely gauge compatible).

From this it would appear that the solutions to our mean-field equations
lead to a large degeneracy of mean-fields. Indeed it would seem that
any gauge transformation of the mean-field solutions leads to a different
solution with the same energy and as such a different ground state.
However, this is not the case: there is only one state of the physical
system that can be obtained from two different states that differ
by a gauge transformation. More precisely, after we project onto the
physical subspace via the projection operators $\Pi_{i}^{S}$, Eq.~\eqref{eq:Constraint-2},
two states that differ by a gauge transformation given in Eq.~\eqref{eq:Gauge_transformation-1}
project onto the same state up to an unobservable overall phase. To
show this, without loss of generality, assume that under the gauge
transformation in Eq.~\eqref{eq:Gauge_transformation-1} the state
with no bosons/fermions transforms into itself, $\left|0\right\rangle \rightarrow\left|0\right\rangle $.
Now an arbitrary state may be written as a linear combination of terms
of the form: 
\begin{equation}
b_{i_{1},\eta_{1}}^{\dagger}\,b_{i_{2},\eta_{2}}^{\dagger}\dots\,b_{i_{n},\eta_{n}}^{\dagger}\:c_{j_{1},\eta_{1},\sigma_{1}}^{\dagger}\dots\,c_{j_{m},\eta_{m},\sigma_{m}}^{\dagger}\left|0\right\rangle \,.\label{eq:State}
\end{equation}
After projection we may as well assume that there are exactly $S$
$b_{i,\eta}/c_{i,\eta,\sigma}$'s at every vertex $i$ in the expression
in Eq.~\eqref{eq:State}. Under a gauge transformation, \begin{widetext}
\begin{eqnarray}
 &  & b_{i_{1},\eta_{1}}^{\dagger}
\dots\,b_{i_{n},\eta_{n}}^{\dagger}c_{j_{1},\eta_{1},\sigma_{1}}^{\dagger}\dots\,c_{j_{m},\eta_{m},\sigma_{m}}^{\dagger}\left|0\right\rangle \rightarrow\nonumber \\
 &  & \!\!\!\!\!\rightarrow e^{-i\theta_{i_{1}}}\,b_{i_{1},\eta_{1}}^{\dagger}\,e^{-i\theta_{i_{1}+\eta_{1}}}
\dots\;e^{-i\theta_{i_{n}}}\,b_{i_{n},\eta_{n}}^{\dagger}\,e^{-i\theta_{i_{n}+\eta_{n}}}\:e^{-i\theta_{j_{1}}}\,c_{j_{1},\eta_{1},\sigma_{1}}^{\dagger}\,e^{-i\theta_{j_{1}+\eta_{1}}}\dots\:e^{-i\theta_{j_{m}}}\,c_{j_{m},\eta_{m},\sigma_{m}}^{\dagger}\,e^{-i\theta_{j_{m}+\eta_{m}}}\left|0\right\rangle \,.\label{eq:Gauge_transformation_state}
\end{eqnarray}
\end{widetext} We can group the phases associated with every vertex
together and, because of the constraint that there are exactly $S$
bosons/fermions at any vertex, this gauge transformation becomes a
state independent phase: 
\begin{equation}
\left\vert \Psi\right\rangle \rightarrow\exp\left(-iS\sum_{i}\theta_{i}\right)\left\vert \Psi\right\rangle \,.\label{eq:Phase_transformation}
\end{equation}

\subsection{\label{sec:The-IGG}The Invariant Gauge Group}

Here we consider the Projective Symmetry Group (PSG) construction,
again focusing on the case where $N=1$. The main idea behind the
PSG (which was originally introduced for spin systems~\cite{key-18,key-19,key-20,key-21,key-22,key-23})
is that in order for a mean-field state to be invariant under a symmetry
transformation of the system (e.g., a translation or a rotation),
the mean-field ansatz needs not remain invariant under the transformation
but needs only be invariant following a gauge transformation of the
form in Eq.~\eqref{eq:Gauge_transformation_hubbard_Stratanovich}
(which does not change the state of the systems as discussed previously).
Using this observation we may define the PSG of a mean-field ansatz~\cite{key-18,key-19,key-20,key-21,key-22,key-23}
as the set of all lattice transformations followed by gauge transformations
(as in Eq.~\eqref{eq:Gauge_transformation_hubbard_Stratanovich})
which leave the mean-field ansatz invariant. An important subgroup
of the PSG is the IGG (Invariant Gauge Group) which is the set of
all gauge transformations in Eq.~\eqref{eq:Gauge_transformation_hubbard_Stratanovich}
that leave the mean field ansatz invariant. We will now proceed to
calculate the IGG for various lattices.

We only focus on HS generated mean-fields of the form in Eq.~\eqref{eq:Mean_field_gauage}
which are gauge transformation compatible via Eq.~\eqref{eq:Gauge_transformation_hubbard_Stratanovich}
and ignore all other mean-fields. It can be checked directly that
the mean-fields obtained by considering the fermion fermion or fermion
boson part of the Hamiltonian do not change the value of the IGG.
We begin with bipartite lattices, with sublattices labeled by $A$
and $B$. The first constraint we obtain by the gauge transformation
in Eq.~\eqref{eq:Gauge_transformation_hubbard_Stratanovich} is that
$x_{i,j}=x_{i,j}$, i.e., it is gauge invariant and therefore it gives
no further restrictions on the form of the gauge transformations.
The second constraint is that 
\begin{align}
w_{i1/2} & =\exp\left[i\left(\theta_{i}-\theta_{i+\hat{x}+\hat{y}}\right)\right]w_{i1/2}\nonumber \\
w_{i3/4} & =\exp\left[i\left(\theta_{i+\hat{x}}-\theta_{i+\hat{y}}\right)\right]w_{i3/4}\,.\label{eq:Constraint_regular}
\end{align}
We assume that the terms $w_{i,j}$ are non-zero (otherwise the mean-field
lattice loses connectivity and breaks up into one dimensional sublattices).
In this case we have that $\exp\left(i\theta_{i}\right)=\exp\left(i\theta_{i+\hat{x}+\hat{y}}\right)\equiv\exp\left(i\theta_{A}\right)$
and $\exp\left(i\theta_{i+\hat{x}}\right)=\exp\left(i\theta_{i+\hat{y}}\right)=\exp\left(i\theta_{B}\right)$.
Therefore the IGG for a bipartite lattice (assuming the terms $z_{i,j}$
are zero) is simply $U\left(1\right)\times U\left(1\right)$ where
there are two different phases living on the two sublattices $A$
and $B$.

When the decoupling fields $z_{i,j}$ are non-zero, we have the constraint
\begin{equation}
\begin{array}{l}
\exp\left[i\left(\theta_{i}+\theta_{i+\hat{x}}+\theta_{i+\hat{y}}+\theta_{i+\hat{x}+\hat{y}}\right)\right]=\\
=\exp\left[2i\left(\theta_{A}+\theta_{B}\right)\right]=1\;,
\end{array}\label{eq:constraint_superconducting}
\end{equation}
so that $\exp\left(i\theta_{A}\right)=\pm\exp\left(-i\theta_{B}\right)$
and the IGG becomes $U\left(1\right)\times\mathbb{Z}_{2}$. Adding
fermions does not change the $U\left(1\right)\times\mathbb{Z}_{2}$.
One simply gets another copy of the same set of equations. Qualitatively
this is because the HS fields transform in a way determined by the
transformation properties of fermion or boson bilinears under Eq.~\eqref{eq:Gauge_transformation-1}.
Since the bosons and fermions transform in the same away under Eq.~\eqref{eq:Gauge_transformation-1},
it does not add any new ``information'' (or constraints) to consider
fermions.

If we consider non-bipartite lattices, the constraint $x_{i,j}=x_{i,j}$
again does not effect the IGG. On the contrary, the constraints in
Eq.~\eqref{eq:Constraint_regular} insure that the phases $\theta_{i}$
and $\theta_{j}$ for any two sites that can be reached by a finite
number of translations along lines joining second nearest neighbors
along a plaquette is the same. Since any two sites on a non-bipartite
lattice may be joined that way, the IGG when the $w_{ij}\neq0$ is
$U\left(1\right)$, i.e., the same phase for every site.

When the terms $z_{ij}$ do not vanish, we have that 
\begin{equation}
\begin{array}{l}
\exp\left[i\left(\theta_{i}+\theta_{i+\hat{x}}+\theta_{i+\hat{y}}+\theta_{i+\hat{x}+\hat{y}}\right)\right]=\exp\left(4i\theta\right)=1\;,\end{array}\label{eq:non_bipartite}
\end{equation}
which means that the IGG is now given by $\mathbb{Z}_{4}$. Adding
fermions does not change the $\mathbb{Z}_{4}$, as once again we get
multiple copies of the same equations. The PSG for the boson/fermion
system may be computed directly. For example one can check that the
Algebraic PSG for the dimer system is the same as the Algebraic PSG
for a bosonic spin liquid on the same lattice with the same IGG~\cite{key-18,key-19,key-20,key-21,key-22,key-23}.
Indeed the gauge degrees of freedom are identical and the ``commutator''
constraint equations~\cite{key-18,key-19,key-20,key-21,key-22,key-23}
are of the same form for the same lattice and the same IGG.


\begin{thebibliography}{10}
\bibitem{key-31} At first glance it appears that there are more gauge
transformations that preserve the Hamiltonian. In particular it looks
like the transformation $b_{i,\eta}\rightarrow e^{i\theta_{i,\eta}}b_{i,\eta}$
and $c_{i,\eta,\sigma}\rightarrow e^{i\theta_{i,\eta}}c_{i,\eta,\sigma}$,
subject to the constraint that the phases around a plaquette add up
to a multiple of $2\pi$ $\theta_{i,\hat{x}}-\theta_{i+\hat{x},\hat{y}}+\theta_{i+\hat{y},\hat{x}}-\theta_{i,\hat{y}}=2\pi n$,
is another set of gauge transformations. Here $n$ is an integer and
$i,\hat{x}$, $i+\hat{x},\hat{y}$, $i+\hat{y},\hat{x}$ and $i,\hat{y}$
are four links arranged counter clockwise around a plaquette. However
one can verify (by explicitly constructing the gauge transformations)
that every gauge transformation of this form can be written as a gauge
transformation of the form in Eq.~\eqref{eq:Gauge_transformation}.
\end{thebibliography}
\end{document}